\documentclass{iopart}

\newcommand{\be}{\begin{eqnarray}}
\newcommand{\ee}{\end{eqnarray}}

\newcommand{\kb}{k_{\rm B}}
\newcommand{\kB}{k_{\rm B}}

\newcommand{\dU}{{\rm d}U}
\newcommand{\dS}{{\rm d}S}
\newcommand{\dV}{{\rm d}V}

\newcommand{\dT}{{\rm d}T}
\newcommand{\dmu}{{\rm d}\mu}

\newcommand{\dQ}{\mbox{d\raisebox{1.3ex}{\rule{-1 ex}{0 em}\rule{1ex}{0.4pt}}}Q}

\newcommand{\dby}[2]{ \frac{{\rm d} #1}{{\rm d} #2}}

\newcommand{\pd}[2]
{ \left( \frac{ \partial {#1}}{\partial {#2}} \right)}
\newcommand{\lpd}[2]
{ ( { \partial {#1}}/{\partial {#2}})}

\usepackage{graphicx}
\usepackage{amssymb}

\usepackage{pgfplots}
\usetikzlibrary{pgfplots.groupplots}
\usetikzlibrary{patterns}

\begin{document}

\title{On determining absolute entropy without quantum theory or the Third Law of thermodynamics}


\author{Andrew M. Steane}
\address{Department of Atomic and Laser Physics, Clarendon Laboratory, Parks Road, Oxford OX1 3PU, England.}


\begin{abstract}
We employ classical thermodynamics to gain information about absolute entropy, without recourse
to statistical methods, quantum mechanics or the Third Law of thermodynamics. 
The Gibbs-Duhem equation yields various simple methods to determine the absolute entropy of a fluid. 
We also study the entropy of an ideal gas and the ionization of a plasma in thermal equilibrium. A single
measurement of the degree of ionization can be used to determine an unknown constant in 
the entropy equation, and thus determine the absolute entropy of a gas. It follows from all these examples
that the value of entropy at absolute zero temperature does not need to be assigned by postulate, but can be
deduced empirically.
\end{abstract}


\maketitle

The aim of this paper is to clarify some issues in classical thermodynamics, concerning
absolute entropy and the Third Law of Thermodynamics. Absolute entropy is simply the total entropy
of a given system, usually indicated by the symbol $S$. In many situations, one does not need to
know the total entropy, only the entropy change $\Delta S$ in some given process. A similar statement applies
to other extensive properties such as mass and volume. However, whereas it is relatively straightforward
to measure total mass or total volume, and to understand the case of zero mass or volume, it is less
straightforward to measure total entropy and to understand the case of zero entropy. For this reason,
the term `absolute entropy' is introduced to draw attention to those cases where we are interested
in the entropy itself, not just the change. The fact that conditions of zero entropy are not self-evident is
also the reason why the Third Law of thermodynamics is introduced. However, this does not mean that we
cannot determine absolute entropy without the Third Law. It is simply that the Third Law is a useful
summary: like any law of physics, it is at once a generalization from empirical observation, and also
part of a self-consistent and elegant theoretical framework.

The Third Law has been useful to physicists in understanding heat capacities and other response functions
at low temperature, and it proved to be very useful and influential
in the foundations of chemistry. It underpins several techniques to evaluate reaction constants and
affinities.

The Third Law, in its modern formulation, asserts that the absolute entropy tends to zero, in the limit
as temperature tends to absolute zero, for each aspect of a system that remains in thermal
equilibrium. The conditions are impossible to realise exactly, in a finite amount of time, but the
statement is very useful nonetheless because in practice many aspects of many systems approach
sufficiently closely to thermal equilibrium at low temperature that their entropy is negligible,
at the lowest achievable temperatures, compared to the entropy at some other temperature of interest.

There are systems, such as a glass, which do not strictly attain a thermal equilibrium
state, so the Third Law as stated above does not apply to them. One may still
associate a temperature with a glass, and make statements about the behaviour of entropy in the low
temperature limit, but the present paper is concerned purely with thermal equilibrium so we do not
need to consider such systems. Similarly, we may for present purposes ignore the case of a degenerate
ground state, by the argument that such a degeneracy would always be lifted in practice by some tiny interaction or
loss of symmetry, so that the true ground state is not degenerate. It is an open question whether that
argument always applies, but in any case, we shall consider systems for which it does apply, and then
comment briefly on the impact of ground state degeneracy at the end.

The introduction of the Third Law into the foundations of classical thermodynamics may leave
one with the impression that, without it, the absolute entropy of a physical system could not be
determined by classical thermodynamic methods, without an appeal to quantum theory. The main
purpose of this paper is to show that this is wrong. But first let us examine the usual account.

The standard
way to determine the absolute entropy at a given temperature $T_{\rm f}$ is to employ an integral such as
\be
S(T_{\rm f}) - S(0) = \int_0^{T_{\rm f}} \frac{ \dQ_{\rm rev} }{T}         \label{SfromintdQ}
\ee
where $\dQ_{\rm rev}$ is the heat entering the system by a reversible process at each temperature $T$.
It is often convenient to use the equivalent expression
\be
S(T_{\rm f}) - S(0) = \int_0^{T_{\rm f}} \frac{ C }{T}  \dT   + \sum_i \frac{L_i}{T_i}  \label{SfromCV}
\ee
where $C(T)$ is the heat capacity under the conditions of the change, 
and $L_i$ are latent heats. If one asserts that $S(0)=0$ on the basis of the Third Law, then
$S(T_{\rm f})$ can be determined by evaluating the integral and sum on the right hand side.
In practice this is done by a combination of empirical observation and mathematical modelling.
The latter is required, for example, in order to carry out an extrapolation to $T=0$.

It is widely believed and asserted that, without the Third Law, classical thermodynamics only makes
statements about entropy changes, and cannot give values of absolute entropy, 
because it is based on equations such as
\be
\dU = T \dS - p \dV.
\ee
We will show that this is not true, in the following sense. Only volume change $\dV$, not absolute
volume, $V$, is involved in the expression for work done on a simple compressible fluid, but it does not
follow from this that there is no ready definition of absolute volume, or empirical method to determine
absolute volume, without appealing to a postulate concerning, for example, the value of $V$ in the limit
of infinite pressure. Similarly, only entropy change $\dS$, not absolute
entropy, $S$, is involved in the expression for heat exchange, but it does not
follow from this that there is no ready definition of absolute entropy, or empirical method to measure it,
without appealing to the Third Law.

We will present several simple empirical methods that allow the absolute entropy of a fluid to be
deduced without any use of quantum theory, and without calorimetry or cooling to near absolute zero. This will clarify what classical thermodynamics can and cannot tell us
about entropy. In particular, it is not true to say that classical thermodynamics can predict entropy changes but not the absolute entropy of an ideal gas, as if entropy were a special case, unlike internal energy, pressure, temperature and so on. In fact in
each of these cases a combination of a modest amount of empirical observation is combined with thermodynamic reasoning in order to yield general information about the system properties, and this is just as true of absolute entropy as it is of other properties.

Of course quantum theory correctly describes the structure of any system, within the limits of our current understanding
of physics, and classical theory does not. An attempt to model a gas in terms of classical particles obeying Newton's laws
of motion, for example, completely fails to describe the low-temperature behaviour correctly. So in this sense quantum
theory is needed and cannot be avoided. However, the argument of the present paper, like many thermodynamic
arguments, is playing a different role. It is showing how thermodynamic reasoning can be invoked to connect one
physical property to another, so that an empirical measurement of one property allows us to infer the other, without the
need for a theoretical model of the structure of the system. That is the sense in which we shall present methods to
determine absolute entropy without the use of quantum theory.

The methods to be discussed do not invoke the Third Law. Therefore,
by also determining the values of the integrals and sum on the right hand side
of (\ref{SfromintdQ}) or (\ref{SfromCV}), one can infer the value of $S(0)$ without appealing
to the Third Law. The Third Law remains a useful observation about the results of such studies.

It follows from this that the value of the entropy at absolute zero cannot be assigned arbitrarily.
It is {\em not} the case that a universal offset in the entropy would have no observable consequences
in classical thermodynamics. The situation is different from that of potential energy, for
example, and gauge freedom more
generally. Some textbooks do not address this issue, but most, such as
Callen \cite{85Callen},
Carrington \cite{94Carrington} or Blundell and Blundell \cite{06Blundell} imply that
the absolute entropy is undetermined without a postulate such as the Third Law,
and some, such as Wilks \cite{61Wilks} and Adkins \cite{83Adkins}, 
include an explicit statement that there is such a freedom, akin to gauge freedom. 
Carrington, p.190 states: ``\ldots it [the Third Law] means that we may consistently fix the
constant of integration in the definition of the entropy of a simple system. \ldots In particular,
we can establish an absolute value for the entropy of each of the chemical species \ldots ." This
is not wrong, but it implies that such fixing and establishing could not be done without the Third Law.
Blundell and Blundell p.193 states: ``Thus it seems that we are only able to learn about {\em changes} in
entropy, \ldots and we are not able to obtain an absolute measurement of the entropy itself." The line of
argument is the same as the one followed by Wilks, p.54, where it is more explicit:
``to obtain absolute values of entropy it is necessary to make use
of the Third Law." Adkins, p.242 states: ``The essential point of the third law is that the constant
is the same for all systems, and it is strictly a matter of convenience to set it equal to zero."
Callen p.279 states: ``In the thermodynamic context there is no a priori meaning to the absolute
value of the entropy." The arguments of the present paper imply that such claims are false.

Section \ref{s.tr} considers the entropy of thermal radiation; section \ref{s.st} the entropy
of an ideal gas, and section \ref{s.general} that of a more general simple compressible system.
All the arguments are related to the Gibbs-Duhem equation, but for the middle section of
the paper that connection is indirect. Sub-section \ref{sub.entropy} uses thermodynamic reasoning to obtain
an expression for the total entropy of an ideal gas whose heat capacity is independent of temperature
and has been given; the expression contains a single unknown additive constant. Sub-section \ref{s.saha}
presents a derivation of an equation equivalent to the Saha equation\cite{20Saha,23Kingdon,08Fridman}, 
which enables the unknown
constant to be determined by the measurement of the degree of ionization of a hydrogen plasma.
Sub-section \ref{s.discuss} discusses the validity of the method and the connection to statistical
methods. The main conclusion of all the examples is that absolute entropy can be determined without
appealing to either a microscopic model or the Third Law.
The paper concludes in section \ref{s.third} by presenting the implications for, and uses of,
the Third Law of thermodynamics, in view of this.

\section{Thermal Radiation}  \label{s.tr}

First, consider thermal radiation. Note, we have no need of a particle
model (or a wave model for that matter) of the radiation inside a cavity in thermal
equilibrium. We require only the knowledge that the energy flux and the pressure
are both directly related to the energy density. Then, the first shows, following
the thermodynamic reasoning of Kirchhoff \cite{83Adkins},
that the energy density $u$ is a function of temperature alone, 
$u=u(T)$ and the second then implies that the pressure also is a function of
temperature alone. In fact one finds $p = u/3$; this can be regarded either as an
empirical observation or as following from classical electromagnetism for isotropic radiation.

The entropy $S$ of a volume $V$ of thermal radiation in a cavity at a temperature $T$ can
be considered as a function of $V$ and $T$, so that we have
\[
\dS = \pd{S}{T}_V\!\dT + \pd{S}{V}_T\!\dV = \pd{S}{T}_V\!\dT + \dby{p}{T} \dV
\]
where we used a Maxwell relation and the fact that $p=p(T)$. Therefore
\be
\pd{S}{V}_T = \dby{p}{T}.  \label{dp}
\ee
But, since $S$ and $V$ are extensive and $T$ is intensive, we have $\lpd{S}{V}_T = S/V$. Using this
in (\ref{dp}) gives
\be
S = V \dby{p}{T}.
\ee
Hence, in order to obtain $S$ for a given $V$, it suffices to measure $p$ as a function of $T$.

One can also, of course, adopt a less direct approach, for example by obtaining
$p=a T^4$ by a standard thermodynamic argument, where $a$ is a constant
related to the Stefan-Boltzmann constant. By measuring the latter, one obtains
$a$ and thus $p(T)$ and hence $S$. This method makes use of the fact 
that the relationship $a = (4/3) \sigma/c$ between $a$ and the Stefan-Boltzmann
constant $\sigma$ 
depends only on energy, not entropy, so does not involve Planck's constant.



We conclude that by measuring either the pressure or the energy flux
at some known temperature, we can find
the absolute entropy of a quantity of
thermal radiation without recourse to the Third Law or quantum theory or any other
model of the microscopic structure of the radiation.

\section{The entropy of an ideal gas}  \label{s.st}

We now turn to the case of an ideal gas. We will describe an experimental method to
determine $S$ which involves
general thermodynamic reasoning and a single measurement of a neutral
hydrogen plasma. The method is quite different to the one invoked for thermal radiation in
the previous section. It relies on a connection between entropy and chemical potential, and requires a greater
amount of theoretical development. First, in sub-section \ref{sub.entropy}, we develop an equation for the entropy of an ideal
gas, which gives the dependence on internal energy, volume and particle number, up to an unknown
additive constant. Then, in sub-section \ref{s.saha}, we show how measurement of a plasma, combined with
the Saha equation, yields the constant. 

\subsection{Entropy as a function of its natural variables}  \label{sub.entropy}

Much of the argument of the present section
is in standard thermodynamics texts, but some texts stop short of the full result.
We begin by considering a closed simple compressible system, and then generalize to open systems. 

For a closed simple compressible system we have that the entropy may be considered a function of two variables, and
we pick temperature and volume for convenience, so that we may write
\be
\dS &=& \pd{S}{T}_V \dT + \pd{S}{V}_T \dV  \nonumber \\
&=& \frac{C_{\rm V}}{T} \dT + \pd{p}{T}_V \dV
\ee
using the definition of constant-volume heat capacity $C_{V}$, and a Maxwell relation. For a system such as
an ideal gas, where the equation of state is $p V = N \kb T$, this gives
\be
S = \int \frac{C_{V}}{T} \dT + N \kb \ln V.
\ee
If we now further assume that the heat capacity is independent of temperature then we have
\be
S = S_0 + C_{V} \ln T + N \kb \ln V,  \label{Sclosed}
\ee
where $S_0$ is an unknown constant. This is a standard `textbook' result\footnote{A general ideal gas
(i.e., a system that obeys Boyle's Law and Joule's Law) need not necessarily have a heat capacity that is independent
of temperature, and in fact many gases that are ideal to good approximation over some range of temperature
have a heat capacity that changes significantly within that range (e.g. nitrogen in the range
300 K--3000 K at one atmosphere). However the restriction to constant heat capacity applies
to any gas over a sufficiently small range of temperature, and 
describes monatomic gases very well at temperatures not near the boiling point.}. 

When the heat capacity is independent of temperature for an ideal gas, one can show that it is given by
\be
C_V = N \kb(\gamma -1)^{-1}  \label{CV}
\ee
where $\gamma$ is the adiabatic index. 

The less widely-known part of the argument generalizes (\ref{Sclosed}) to open systems by obtaining
the dependence of $S_0$ on $N$. The method is to use the extensive nature of entropy. 
If one begins the argument afresh, but treating $S=S(T,V,N)$, then one obtains  (\ref{Sclosed}) again
but now $S_0$ is a function of $N$, not a constant. To find the dependence, one may use
the fact that
$S(\lambda U, \lambda V, \lambda N) = \lambda S(U,V,N)$ for any $\lambda$, and therefore
$S(U,V,N) = (N/N_0) S(N_0 U/N,\, N_0 V/N,\, N_0)$ by choosing $\lambda = N_0/N$.
To make use of this, one needs also the relationship between internal energy and temperature, which is
\be
U = U_0 + C_V T  \label{UT}
\ee
where $U_0$ is a constant of integration. Empirically, this $U_0$ is the intercept with the $U$ axis
of a line tangential to $U(T)$, but note, we do not require that (\ref{UT}) is valid all the way to
absolute zero, and in fact it is not, because the gas will liquify and/or solidify and
enter the quantum degenerate regime at low temperatures. We have
included $U_0$ here (often it is set equal to zero) in
order to be clear about what assumptions we do and do not need to make in the argument. Since $U$ and
$C_V$ are extensive, and $T$ is not, it is clear that this $U_0$ must also be extensive, which we use
in the following.

After expressing the temperature in terms of energy, and using the extensive nature of entropy,
the result is
\be
S(U,V,N) = N k_B \ln \left[ \zeta \frac{V}{N} \left( \frac{U-U_0}{N} \right)^{\displaystyle \frac{1}{\gamma-1}} \right] ,
\label{entropy}
\ee
where $\zeta$ is a constant that does not depend on $U,\,V$ or $N$. This is the equation giving the
entropy of an ideal gas of constant heat capacity, in terms of the natural
variables of $S$. The complete thermodynamic behaviour can be derived from it. When the value of
the constant $\zeta$ is also given in terms of fundamental quantities, this is the Sackur-Tetrode equation.
Our interest here is in obtaining $\zeta$ from empirical measurements.

\subsection{Saha equation}  \label{s.saha}

We now turn to the consideration of the thermal equilibrium of a hydrogen gas. Such a gas is ordinarily
considered to consist of a set of molecules, or, when dissociated, neutral atoms, in
thermal motion, and this is a very
good approximation for a wide range of densities and temperatures. However, in fact there is
always some non-zero degree of ionization, and this becomes especially important in the context
of astrophysics and plasma physics, where it is a major source of information about the photospheres
of stars and other plasmas. In the following we consider an atomic hydrogen gas, allowing for
its non-zero state of ionization in thermal equilibrium, but in conditions where the degree of
ionization is small.

We recall the standard derivation of the Saha equation\cite{20Saha,23Kingdon,08Fridman} which
gives the degree of ionization
in terms of temperature and atomic properties. The ionization/recombination process is
\[
{\rm H} \leftrightharpoons {\rm p}^{+} + {\rm e}^-
\]
and at equilibrium one has
\be
\mu_{\rm H} = \mu_{\rm p} + \mu_{\rm e}   \label{Sahaequil}
\ee
where the subscripts refer to hydrogen atoms, protons and electrons, respectively.
This condition can be expressed in terms of other quantities by treating the protons,
electrons and hydrogen atoms as a mixture of ideal gases, but we must account correctly for
the binding energy $E_R$. The effect of the latter can be understood as follows.
Each gas in the mixture obeys eqn (\ref{UT}), so we have
\be
U_i = U_{0,i} + C_{V,i} T.
\ee
where the subscript $i$ runs over the proton, electron and neutral hydrogen gas.
If one were studying any one gas in isolation, then the value of $U_{0,i}$ for that gas
could be set arbitrarily, but when the three interact, the values of $U_{0,i}$ are not
independent. Since each $U_i$ is extensive, we have
\be
U_{0,i} = N_i u_i
\ee
where $u_i$ is that part of the energy per particle that does not depend on temperature.
By conservation of energy, we have
\be
u_{\rm H} +  E_R = u_{\rm p} + u_{\rm i} .               \label{energybalance}
\ee
where $E_R$ is the energy required to ionize a hydrogen atom. In the
following we will take it that this binding
energy is given, to good approximation, by the ground state ionization energy, 
also called Rydberg energy.
This approximation ignores the internal thermal excitation of the hydrogen atoms; it assumes
that in order to ionize an atom the full Rydberg energy must be provided. This is a good approximation
for $\kb T \ll E_{\rm R}$; we assume this limit for the rest of the argument.
Since $E_{\rm R} / \kb \simeq 1.5 \times 10^5\,$K,
the approximation will hold very well in typical laboratory conditions (as well as in the photospheres
of stars). Note, also, that the entrance of $E_R$ into the argument does not imply that we have assumed
any particular model of hydrogen atoms, such as the one provided by quantum physics. We require
only the empirical observation that a fixed amount of energy is required to ionize each atom.

In a standard approach, one might now introduce a formula such as
$ \mu = \kb T \ln (n \lambda_T^3)$
where $\lambda_T$ is the thermal de Broglie wavelength. This formula can be obtained, for example,
by a statistical mechanical argument. Here we will not need that argument, because we can instead obtain
an expression for chemical potential from the entropy. 

Consider the Gibbs function of a single-component ideal gas:
\be
G = \mu N = U - TS + p V.             \label{Gibbs}
\ee
For a gas having constant heat capacity we may use (\ref{UT}) and therefore
\be
\mu  = \left(\kb+ \frac{C_V}{N}\right) T - T \frac{S}{N}  + \frac{U_0}{N} .  \label{muS}
\ee
We now apply this equation separately to the electrons, protons and hydrogen atoms in the plasma. This
is justified by the usual argument that, at low density, different gases in a mix of ideal gases contribute
their entropies independently, and here it also involves an assumption of overall neutrality of the plasma,
in order to justify ignoring the contribution of the electromagnetic field. Under the approximation that
the latter can be neglected, we may substitute (\ref{muS}) three times into (\ref{Sahaequil}) 
and thus obtain
\be
T \left( s_{\rm e} + s_{\rm p} - s_{\rm H}  \right) =  \frac{5}{2} \kb T + \alpha
\ee
where we used $C_V/N = (3/2) \kB$ for each species, $s_i$ is the entropy per particle
in each case, and 
\be
\alpha = u_{\rm p} + u _{\rm e} - u_{\rm H} = E_R           \label{alpha}
\ee
Using now the entropy equation, (\ref{entropy}), to express $s_{\rm H}$ and $s_{\rm p}$ 
in terms of the constants $\zeta_{\rm H}$, $\zeta_{\rm p}$ and other quantities, we find
\be
s_{\rm e} = \kb \left[ \frac{5}{2} + \frac{E_{\rm R} }{\kb T} 
+ \ln \left( \frac{\zeta_{\rm H} n_{\rm p}}{\zeta_{\rm p} n_{\rm H}} \right) \right]     \label{selectron}
\ee
where $n_i$ is number density. 

An alternative derivation of eqn (\ref{selectron}) is given in the appendix.

Next, we argue that $\zeta_{\rm H} \simeq 2 \zeta_{\rm p}$, to an accuracy
of order one part per thousand, because when the effect of electrical charge is neutralized by the presence
of the electrons, the gas of hydrogen atoms and the gas of protons are alike, except
one has four energetically available
internal states per particle, the other two. This neglects the slight difference between the mass
of a neutral hydrogen atom and the mass of a proton. The factor 2 can be argued as a
thermodynamic statement, not an appeal to Boltzmann's formula $S = \kB \ln W$,
because when all the spin states are equally occupied one can always model
each gas as an equal mixture of distinguishable gases, one in each spin state, and
apply the entropy formula to each part in the mixture. This point is presented more fully
in the appendix.
Hence we find that,
under the stated approximations, the entropy of a gas of electrons forming a component of a
neutral hydrogen plasma in thermal equilibrium is given by
\be
S_{\rm e} = N_{\rm e} \kb \left[ \frac{5}{2} + \frac{E_{\rm R}}{\kb T} 
+ \ln \left( \frac{2 n_{\rm p}}{n_{\rm H}} \right) \right]          \label{Selectron}
\ee
where $N_{\rm e}$ is the number of electrons, $E_{\rm R}$ is the Rydberg energy
and $n_{\rm p}$, $n_{\rm H}$ are the number densities of protons and hydrogen atoms,
respectively, in the plasma. Equation (\ref{Selectron}) may be regarded as
a way of writing the Saha equation that is convenient to our purposes. The significance of
our analysis is that it shows how the properties of the plasma may be related directly to
entropy using only classical thermodynamic reasoning, without the use of statistical
mechanics or quantum theory.

The experimental determination of $S_{\rm e}$ now proceeds by measuring 
the density and the degree of ionization at a known
temperature, and deducing the entropy of the electron gas from (\ref{Selectron}). 
This is the total entropy, without unknown additive constants, sometimes called the
absolute entropy. By using this result to supply the value of
the otherwise unknown constant $\zeta$ in equation (\ref{entropy}), one obtains the entropy under
all conditions for an ideal gas of electrons in a neutral plasma.  Eqns (\ref{Selectron}) and
(\ref{entropy}) can be combined to give an explicit formula for $\zeta_{\rm e}$ in terms of the
measured quantities:
\be
\zeta_{\rm e} = 2 e^{5/2} \frac{n_{\rm p} n_{\rm e}}{n_{\rm H}} \left( \frac{3}{2} \kb T \right)^{-3/2}
e^{E_{\rm R}/\kb T}.   \label{ae}
\ee
(In a neutral plasma, $n_{\rm e} = n_{\rm p}$ and $n_{\rm H} + n_{\rm p} = $ constant.)


\subsection{Discussion}  \label{s.discuss}

Eqns (\ref{ae}) and (\ref{entropy}) give our second example of 
the main point asserted in this paper. They show that, in order to
find out how much entropy a monatomic gas has got, under ordinary conditions where it
behaves to good approximation like an ideal gas, one does not need quantum theory and
one does not need to carefully track the heat supplied as the system is warmed reversibly from
absolute zero. This is because the law of mass action (of which the Saha equation is an example) can be
converted into a statement about entropy, instead of the usual form in terms of chemical potential.
This observation may not be completely new, since similar reasoning applies to a large
number of chemical reactions and chemists are familiar with this type of strategy for
avoiding calorimetry when studying chemical potential. However, we have brought out some
implications for the Third Law that appear to have been overlooked.

The thermodynamic point here is that the calculation of the absolute entropy does require
empirical input, of course it does---because one cannot say anything about a system
which has not been measured or specified in some way---but the amount of empirical
input is small compared to the amount of physical prediction that can then be deduced. This is similar
to the observation of other properties such as internal energy and temperature.

We presented the argument directly in terms of entropy, in order to arrive at eqn (\ref{Selectron}),
but some readers may find the following approach more intuitive. First we write down a
fundamental equation relating to chemical potential of a single component system:
\be
\mu = \pd{U}{T}_{T,V} - T \pd{S}{N}_{T,V}.
\ee
Next we use (\ref{UT}) and (\ref{entropy}), to obtain
\be
\mu = \frac{U_0}{N} + \kB T \left( \frac{\gamma}{\gamma-1} - \ln \zeta -
 \ln\left[ \frac{1}{n} \left( \frac{\kb T}{\gamma-1} \right)^{1/(\gamma-1)}  \right]\right).
\label{muandzeta}
\ee
This draws attention to the fact that $\zeta$ appears as an unknown offset in the 
equation for $\mu$, just as it does in the equation for $S$. The essence of the argument is to see
that we can obtain this unknown offset by using the fact that in chemical equilibrium
the chemical potentials balance (eqn (\ref{Sahaequil})), in a situation where 
the contribution of potential energy to the energy per particle is known, eqn (\ref{energybalance}),
and this contribution has to be made up by the translational degrees of freedom
in the chemical potential balance.
By substituing (\ref{muandzeta}) three times into (\ref{Sahaequil}) and employing
(\ref{energybalance}), one obtains (\ref{ae}) as before.



\begin{figure}
\begin{tikzpicture}[scale=0.8]
\begin{axis}[
xmin=0, xmax=80,
ymin=-10, ymax=130,
xlabel={$T (K)$},
ylabel={$S (N_{\rm A}k_{\rm B})$},
]
\addplot [ ]
coordinates {
(0,0)
(80,0)
};
\addplot [ ]
coordinates {
(0,0)
(0.501224,0.00590381)
(1.00245,0.0236152)
(1.50367,0.0531343)
(2.0049,0.094461)
(2.50612,0.147595)
(3.00735,0.212537)
(3.50857,0.289287)
(4.0098,0.377844)
(4.51102,0.478209)
(5.01224,0.590381)
(5.51347,0.714361)
(6.01469,0.850149)
(6.51592,0.997744)
(7.01714,1.15715)
(7.51837,1.32836)
(8.01959,1.51138)
(8.52082,1.7062)
(9.02204,1.91283)
(9.52327,2.13128)
(10.0245,2.36152)
(10.5257,2.60358)
(11.0269,2.85744)
(11.5282,3.12312)
(12.0294,3.40059)
(12.5306,3.68988)
(13.0318,3.99098)
(13.5331,4.30388)
(14.0343,4.62859)
(14.5355,4.9651)
(15.0367,5.31343)
(15.538,5.67356)
(16.0392,6.0455)
(16.5404,6.42925)
(17.0416,6.82481)
(17.5429,7.23217)
(18.0441,7.65134)
(18.5453,8.08232)
(19.0465,8.5251)
(19.5478,8.9797)
(20.049,9.4461)
(20.5502,9.92431)
(21.0514,10.4143)
(21.5527,10.9161)
(22.0539,11.4298)
(22.5551,11.9552)
(23.0563,12.4925)
(23.5576,13.0415)
(24.0588,13.6024)
(24.56,14.175)
(24.56,41.4552)
(27.104,46.9257)
(27.104,96.376)
(28.928,97.7298)
(30.752,99.0007)
(32.576,100.198)
(34.4,101.331)
(36.224,102.405)
(38.048,103.426)
(39.872,104.399)
(41.696,105.329)
(43.52,106.219)
(45.344,107.072)
(47.168,107.892)
(48.992,108.681)
(50.816,109.441)
(52.64,110.174)
(54.464,110.882)
(56.288,111.567)
(58.112,112.229)
(59.936,112.872)
(61.76,113.495)
(63.584,114.1)
(65.408,114.688)
(67.232,115.26)
(69.056,115.816)
(70.88,116.358)
(72.704,116.886)
(74.528,117.401)
(76.352,117.904)
(78.176,118.394)
(80,118.874)
};
\addplot [dashed]
coordinates {
(0.05,-34.4816)
(0.352525,6.11585)
(0.655051,18.9948)
(0.957576,26.8871)
(1.2601,32.5938)
(1.56263,37.0665)
(1.86515,40.7451)
(2.16768,43.8696)
(2.4702,46.5851)
(2.77273,48.9866)
(3.07525,51.1391)
(3.37778,53.0895)
(3.6803,54.8725)
(3.98283,56.5145)
(4.28535,58.0363)
(4.58788,59.4542)
(4.8904,60.7816)
(5.19293,62.0292)
(5.49545,63.2062)
(5.79798,64.3201)
(6.10051,65.3773)
(6.40303,66.3834)
(6.70556,67.343)
(7.00808,68.2602)
(7.31061,69.1387)
(7.61313,69.9815)
(7.91566,70.7915)
(8.21818,71.5712)
(8.52071,72.3226)
(8.82323,73.0478)
(9.12576,73.7486)
(9.42828,74.4265)
(9.73081,75.0829)
(10.0333,75.7193)
(10.3359,76.3368)
(10.6384,76.9365)
(10.9409,77.5193)
(11.2434,78.0863)
(11.546,78.6382)
(11.8485,79.1758)
(12.151,79.6999)
(12.4535,80.2111)
(12.7561,80.71)
(13.0586,81.1972)
(13.3611,81.6732)
(13.6636,82.1386)
(13.9662,82.5938)
(14.2687,83.0393)
(14.5712,83.4754)
(14.8737,83.9025)
(15.1763,84.3211)
(15.4788,84.7313)
(15.7813,85.1337)
(16.0838,85.5284)
(16.3864,85.9157)
(16.6889,86.296)
(16.9914,86.6694)
(17.2939,87.0362)
(17.5965,87.3967)
(17.899,87.751)
(18.2015,88.0994)
(18.504,88.4421)
(18.8066,88.7791)
(19.1091,89.1109)
(19.4116,89.4374)
(19.7141,89.7588)
(20.0167,90.0754)
(20.3192,90.3872)
(20.6217,90.6944)
(20.9242,90.9971)
(21.2268,91.2955)
(21.5293,91.5896)
(21.8318,91.8797)
(22.1343,92.1657)
(22.4369,92.4479)
(22.7394,92.7263)
(23.0419,93.001)
(23.3444,93.2721)
(23.647,93.5398)
(23.9495,93.804)
(24.252,94.0649)
(24.5545,94.3226)
(24.8571,94.5772)
(25.1596,94.8286)
(25.4621,95.0771)
(25.7646,95.3226)
(26.0672,95.5652)
(26.3697,95.8051)
(26.6722,96.0422)
(26.9747,96.2766)
(27.2773,96.5084)
(27.5798,96.7377)
(27.8823,96.9645)
(28.1848,97.1888)
(28.4874,97.4107)
(28.7899,97.6303)
(29.0924,97.8476)
(29.3949,98.0626)
(29.6975,98.2754)
(30,98.4861)
};
\end{axis}
\end{tikzpicture}
\caption{The entropy of one mole of isotopically pure neon at one atmosphere, as a function of temperature.
The full line is the experimental result, the dashed line is the prediction of eqn (\protect\ref{Stotquant}).
The latter gives unphysical negative values as $T \rightarrow 0$, but it 
describes the gaseous behaviour very well. (The empirical curve was obtained by combining information
about heat capacities and latent heat.)}
\label{f.Sgas}
\end{figure}
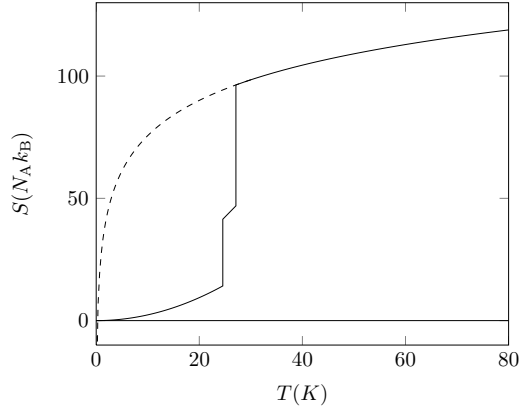

The fact that no gas is ideal at low temperatures does not invalidate the 
method for measuring entropy that we have described. Indeed, equation (\ref{entropy}) is 
completely wrong at low temperatures; this does not invalidate the argument but rather serves
to emphasize that we have no need of the Third Law or heating from absolute zero
in order to arrive at our result.
Equation (\ref{entropy}) is correct in its regime of validity
(namely, low density, and over modest changes in temperature), and the observation of the
ionization fraction will determine the
constant $\zeta$ correctly.  To illustrate this point, figure \ref{f.Sgas} shows the entropy of
one mole of neon at one atmosphere, as a function of temperature. When we model
neon as an ideal gas, we do not learn about its entropy at or below the boiling
point, but this does not mean that our equation for the absolute entropy (not just entropy
changes) lacks accuracy above the boiling point. This can be clarified by writing
eqn (\ref{selectron}) another way:
\be
\frac{n_{\rm p}}{n_{\rm H}} = \frac{1}{2 e^{5/2}} e^{s_{\rm e}/\kb} e^{-E_{\rm R}/\kb T}
\ee
With the benefit of statistical reasoning, we can interpret the right hand side of this formula
as the product of a number of states and a Boltzmann factor. The observation of the ionization
fraction enables one to learn what the number of states is, when the energy $E_{\rm R}$ and
the temperature are known. 

We will now present the connection to statistical mechanics more fully. According to classical statistical
mechanics, the entropy is given by Boltzmann's formula $S = \kb \ln W$, and $W$,
the number of states available to the system, is counted by dividing the accessible phase
space volume per particle by some constant $h^3$ whose value has to be determined empirically. Of course
quantum theory comes in and tells us that this constant $h$ is none other than Planck's constant that
relates energy to frequency and momentum to wavelength of de Broglie waves, but one
does not need to have that further information in order to make use of classical statisitical
mechanics combined with the empirically determined unit of volume in phase space.
Whether a classical or
a quantum treatment is adopted, one finds that in the case of a monatomic
gas at low pressure (where particle interactions can be neglected), the entropy is
\be
S = N\kb \left( \frac{5}{2} - \ln n \lambda_T^3 \right)  \label{Stotquant}
\ee
where $\lambda_T$ is the thermal de Broglie wavelength given by
\be
\lambda_T = \sqrt{ \frac{2 \pi \hbar^2}{m \kb T} }.
\ee
This implies that for the monatomic gas, the constant $\zeta$ in (\ref{entropy}) is
$\zeta = e^{5/2} (3\pi\hbar^2/m)^{-3/2}$, where $m$ is the mass of one atom.
Hence the measurement of $\zeta$ can be regarded either as a way to obtain the entropy, or as a
way to determine $\hbar$. Once this has been done in one case (such as the electron gas), one can
then use the formula to find the entropy of any other ideal gas. 
In order to compare this result with eqn (\ref{ae}) one should note
that (\ref{ae}) includes the contribution from the spin degree of freedom for a gas of electrons,
whereas (\ref{Stotquant}) applies to a gas of spinless particles.

As an overall consistency check, we quote here the standard statement of the Saha equation in
the case of hydrogen:
\be
\frac{n_{\rm p} n_{\rm e}}{n_{\rm H}} = \left( \frac{m_{\rm e} k_{\rm B} T}{2 \pi \hbar^2} \right)^{3/2}
e^{-E_R/\kb T}.
\ee
If one substitutes this into the right hand side of (\ref{ae}) one finds $\zeta_{\rm e}
= 2 e^{5/2}  (3\pi\hbar^2/m_{\rm e})^{-3/2}$ which is the Sackur-Tetrode result for a gas
of spin-half particles.




\section{A more general system}  \label{s.general}

We turn now to the case of a sample of ordinary matter (not radiation, and not necessarily 
an ideal gas).
We will restrict our attention to a sample that can be treated as
an ideal fluid, i.e., a system that exhibits pressure but not sheer stress, but the sample
need not be a gas and may have any equation of state. 

Our proposed method is based
on the Gibbs-Duhem relation. For an ideal fluid the Gibbs-Duhem relation reads
\be
S \dT - V {\rm d}p + \sum_i N_i \dmu_i  = 0
\ee
where the symbols have their usual meanings. $N_i$ is the number of particles of
the $i$'th component, $\mu_i$ is the chemical potential per particle. In the case of
a single component system, we have
\be
S = V \pd{p}{T}_\mu          \label{SfromGD}
\ee
and
\be
S = N \pd{\mu}{T}_p  .        \label{SfromGDp}
\ee
Both of these equations offer ways to find $S$. The treatment of thermal radiation in
section \ref{s.tr} can be regarded as an example of (\ref{SfromGD}) if one models
thermal radiation as a fluid at zero chemical potential. The argument from the Saha equation 
presented in section \ref{s.saha} could be regarded as indirectly or loosely connected to 
eqn (\ref{SfromGDp}).

We now present a further application of eqn (\ref{SfromGD}). 

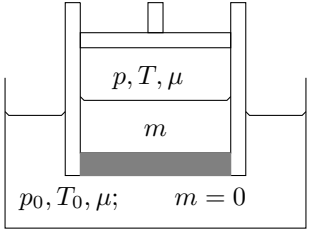
\begin{figure}
\begin{tikzpicture} [scale=1]
\draw (0,2) -- (0,0) -- (4,0) -- (4,2);
\draw (1,3) -- (1,1) -- (3,1) -- (3,3) -- (3.2,3) -- (3.2,0.7) -- (0.8,0.7) -- (0.8,3) -- (1,3);
\draw[pattern=dots,color=gray] (1,0.7) rectangle (3,1);
\draw (1,2.4) rectangle (3,2.6);
\draw (1.9,2.6) rectangle (2.1,3);
\draw (1,1.75) -- (1.05,1.7) -- (2.95,1.7) -- (3,1.75);
\draw (0,1.55) -- (0.05,1.5) -- (0.75,1.5) -- (0.8,1.55);
\draw (3.2,1.55) -- (3.25,1.5) -- (3.95,1.5) -- (4,1.55);
\draw (1.9,2) node {$p,T,\mu$};
\draw (2,1.3) node {$m$};
\draw (1.7,0.4) node {$p_0,T_0,\mu; \;\;\;\;\;\;\; m=0$};
\end{tikzpicture}
\caption{Experimental method to find the entropy of a nearly-pure vapour. Here $m$ is the
concentration of a solute in a solvent and the barrier between the two chambers is a semipermeable channel
that allows
solvent but not solute to pass. Semipermeable membranes on each side of the channel
support a pressure difference, and the channel is
thermally insulating to good approximation (see main text).
There is phase equilibrium between the solution and its vapour
in the inner chamber. The fixed $p_0,\,T_0$ ensure that $\mu$ is constant in the outer chamber.
By allowing the concentration $m$ in the inner chamber to change, one adjusts the position of the phase equilibrium
line as a function of pressure and temperature. This allows one to explore a range of $p,\,T$ values
all at the same $\mu$. When the vapour pressure of the solute is low, the vapour is almost a pure substance,
so one learns the entropy of that pure substance in its vapour phase. The vapour may have any
equation of state.}
\label{f.expt2}
\end{figure}

We can apply  (\ref{SfromGD})
straightforwardly to any experiment in which the pressure and temperature of a fluid are
caused to vary quasistatically, under conditions of constant chemical potential. Such conditions
are not achieved under ordinary circumstances. An idealized particle reservoir can provide
particles without also providing entropy, and such a reservoir is not ruled out
in principle \cite{85Waldram}, but what one wants is a practical method. Figure \ref{f.expt2} illustrates
the main components of such a method. 

A chamber of adjustable pressure and temperature $p,T$ is surrounded by another whose pressure
and temperature $p_0,T_0$ are maintained constant. 
The wall between the two chambers consists, in whole or in part,
of a section that is permeable to a solvent such as water, but not to a solute such
as common salt (sodium chloride). This section could consist, for example, of a short channel
covered by semi-permeable membranes. The inner chamber contains a pool of 
solvent with solute dissolved, in phase equilibrium with its vapour. 
In this circumstance, solvent will
move between the chambers until the chemical potential $\mu$ of the solvent is the same in two chambers,
and at equilibrium there will be a uniform temperature $T = T_0$, and a 
pressure difference owing to osmotic pressure, such that $p > p_0$.

We now change the concentration $m$ and the temperature $T$ in the inner
chamber, by small amounts $\delta m,\delta T$,
while keeping $p_0$ and $T_0$ fixed in the outer chamber. Owing to the fixed $p_0,T_0$, the 
solvent's chemical potential
$\mu$ does not change in the outer chamber. Since the semi-permeable membranes allow solvent
to pass, the condition of chemical equilibrium will be attained when $\mu$ 
is also unchanged in
the inner chamber. Once this has happened, there will be a new value of the osmotic pressure,
so $p$ will have changed by some small amount $\delta p$. One then has a change in both $p$ and $T$
without a change in $\mu$, so by measuring both for a few values and taking the limit, one obtains,
subject to a {\em proviso} presented next, an empirical value for $\lpd{p}{T}_\mu$ for the solvent in this experiment. 

Before we can make such a claim, the following consideration has to be taken into account.
At the new values $p +\delta p$ and $T + \delta T$, the fluid in the inner chamber will not
be in a strict thermal equilibrium, because of heat transport to the outer chamber. One would like
the semipermeable channel to be also thermally insulating, but this is not strictly possible,
owing to heat transport by convection as the solvent passes between the chambers.
However, one can have a case where there is a large ratio between the relevant relaxation times.
There are three relaxation times to be considered. Let $\tau_\mu$ be the time required
for the equality of the chemical potential to be established, $\tau_T$ be the time required for
the fluid in the inner chamber to attain its own thermal equilibrium if it were in complete thermal
isolation, and $\tau_{\rm c}$ be the time for temperature equilibrium to be established between
the chambers in the actual system. If the channel has a low
coefficient of thermal transport (including the contributions from both convection and ordinary thermal
conduction) then one can have
\be
\{ \tau_\mu,\; \tau_T \} \ll \tau_{\rm c} .      \label{taus}
\ee
In this situation, the conditions in the inner chamber are close to thermal equilibrium conditions
at the new temperature $T +\delta T$ for times $t$ satisfying $\{ \tau_\mu,\;\tau_T \} \ll t \ll  \tau_{\rm c}$.

We have now established that, as long as the condition (\ref{taus}) is satisfied, then the measurements
of $\delta T$ and $\delta p$ can be interpreted in the ordinary way, and one can thus obtain an accurate
measurement of  $\lpd{p}{T}_\mu$. It only remains to comment on the role of the vapour in
the inner chamber. The chemical solution in the inner chamber is a system of two chemical components, and
so equation (\ref{SfromGD}) does not apply to it. Instead another equation applies, involving
the chemical potentials of both the solvent and the solute. However, it is quite common for
the vapour pressure of a solute to be very much lower than that of a solvent, so that the
vapour consists almost entirely of the solvent. In this case the vapour is a one-component
system to which  (\ref{SfromGD}) applies, and thus the experiment yields the absolute entropy
of a given volume of the vapour (where we have used the fact that, in phase equilibrium,
$\mu$ is the same for the two phases).

Note that although we assumed the vapour did not exhibit sheer stress, we did not need to assume
anything about its equation of state. In short, it need not be an ideal gas. Also, although we have
called it a vapour in the above, the second phase of the matter in the inner chamber need not
be a gaseous phase; we only require that it be a phase to which the single-component
Gibbs-Duhem equation can be applied to good approximation. This includes some solids as well
as many liquids and gases.

The experimental method we have described is very closely related to
the celebrated `fountain effect' in liquid helium.
Here, two chambers containing liquid helium below the lambda point are separated by a 
narrow tube or plug
which is porous to the superfluid but not the ordinary component of the liquid. In the
presence of a temperature difference between the chambers, superfluid moves from the
colder to the hotter chamber until a pressure difference builds up, given by equation 
(\ref{SfromGD}). One can use the equation in two ways. If the entropy is obtained by using
the Third Law and integrating the heat capacity, one predicts the pressure difference for
a given temperature difference. Alternatively, one may measure the pressure difference, and
use it to obtain the entropy per unit volume of the liquid helium.

\section{The Third Law} \label{s.third}

The Third Law of thermodynamics makes a statement about the absolute value of the entropy
of any physical system, as explained in the introduction. 
The usual application of the Third Law is to say that since $S(0)=0$, the right hand side
of expressions (\ref{SfromintdQ}) and (\ref{SfromCV}) can be used to determine $S(T_{\rm f})$. Expressions 
(\ref{entropy}), (\ref{ae})
and 
(\ref{SfromGD}) offer an alternative application. If, by using those equations, we learn $S(T_{\rm f})$
for a given system such as a mole of neon at some temperature, and then we use the
Third Law to claim that $S(0) = 0$, then we obtain a {\em prediction} for the value of the right
hand side of eqns (\ref{SfromintdQ}) and (\ref{SfromCV}). Note, this prediction is not
about the gas alone, but about an integral involving the heat capacity of a solid and a liquid and a gas, and
a sum involving the values of various latent heats and transition temperatures. 

An alternative application of our results is to ignore the Third Law, and use 
(\ref{SfromintdQ}) and (\ref{SfromCV}) to determine the value of $S(0)$.
For many systems, one will find the value $S(0) = 0$, and one will never
find a negative value. It is not necessary to invoke either classical or quantum statistical mechanics
here; one simply makes the measurement. Of course, in this approach one will eventually
formulate the Third Law as a useful summary of what is found in such experiments.
It can be stated either as we have done, in terms of thermal equilibrium and the
value $S(0) = 0$, or one may invoke quantum theory and discuss the ground state
and its possible degeneracy. In either case one finds that, for any given system, the entropy in the limit
$T \rightarrow 0$ takes a single value that does not change in isothermal processes,
and one can use the results of this paper to discover what that value is. The point is,
the value is fixed absolutely by a measurement at high temperature and
the use of calorimetry to track the entropy reduction as the temperature falls.

It is sometimes asserted
that the value of the constant towards which the entropy tends as $T \rightarrow 0$ is not
fixed by thermodynamic arguments, and could in principle be some other value, as long as it is
the same for all aspects of all systems. Such an assertion is in Adkins' book, for example \cite{83Adkins}.
Callen \cite{85Callen} makes a similar statement.
One of the important features of the argument of this paper is that it shows that such is not
the case. The value of $S(T)$ as $T \rightarrow 0$ is not arbitrary but 
can be ascertained  by measurement, without requiring statistical or quantum theory. 
This does not mean the Third Law is not useful, but it does mean
that the value of the entropy at absolute zero is a measurable, not an arbitrarily assigned, quantity,
within classical thermodynamics.


All such arguments are not necessary to physics if one assumes that some fundamental theory, such as
quantum theory and the Standard Model, is correct, and if one also assumes that we know how to apply 
such a model in order to gain understanding. 
However, thermodynamic arguments remain important because their role is, in part, to show us 
what must be true of the physical world, under a small number of assumptions, irrespective
of our microscopic models. Their role is also, in part, to show us what concepts and ways of reasoning
are fruitful. 

I thank Stephen Blundell and two anonymous referees for helpful comments on the paper.

\section{Appendix}


{\em Alternative derivation of eqn (\ref{selectron}).}

Textbook derivations of the Saha equation often adopt a method slightly different to the
one presented in section \ref{s.saha}, as follows. 

Consider first the definition of chemical potential,
\be
\mu_i \equiv \pd{U}{N_i}_{S,V,\{N_j\}}.   \label{muidef}
\ee
For a gas, the internal energy can be conveniently divided into the kinetic energy
associated with translational motion of the particles, and the rest (potential
energy and the energy in internal degrees of freedom of the particles themselves).
Since the internal energy can thus be regarded as a sum of two parts, so can the
chemical potential. Let us attach a star to the part associated purely with translational
kinetic energy, giving
\be
\mu_i^* \equiv \pd{U^*}{N_i}_{S,V,\{N_j\}}.   \label{mustar}
\ee
Then for a case where the internal degrees of freedom contribute
negligibly to the heat capacity, we have $U^* = \sum_i U_i^*$ where
\be
U_i^* = C_{V,i} T .
\ee

In the neutral hydrogen plasma,
the energy required to
remove one hydrogen atom and provide one proton and one electron, without
changing the system entropy, is
\be
-\mu_{\rm H}^* + \mu_{\rm p}^* + \mu_{\rm e}^* + E_R . \label{muSahastar}
\ee
The symbol $\mu_{\rm H}^*$ here can be interpreted as the energy liberated when a hydrogen
atom in its ground state is removed from the mix without first 
exciting it with the energy $E_R$ that would be required for ionization. 
The binding energy $E_R$ is normally associated with
the hydrogen atom in eqn (\ref{muSahastar}), so one defines
$\mu_{\rm H} = \mu_{\rm H}^* - E_R$
but this is not the only way to understand the situation: the energy $E_R$ is an interaction
energy so it really belongs to the electromagnetic field. For the purposes of book-keeping,
it could be assigned to any one of the particles, or shared among them. The important point is to write down
correctly what is the total energy conservation condition, allowing for all the energies involved.
This is done by asserting that, in equilibrium, the chemical potentials are so arranged that
the sum in (\ref{muSahastar}) is zero. That is,
\be
\mu_{\rm p}^* + \mu_{\rm e}^* + E_R   - \mu_{\rm H}^* = 0.      \label{muERstar}
\ee
This is another way of writing eqn (\ref{Sahaequil}).
Note that we have carefully avoided here any statements about partition functions. Indeed, the
fact that the partition function for the electronic excitations of 
a single hydrogen atom in an infinite volume is itself infinite is
commonly ignored or glossed-over in textbook treatments\footnote{For a single
hydrogen atom in an infinite volume, the partition function associated purely with the
electronic excitations, even without ionization, is
$Z = \sum_{n=1}^{\infty} 2 n^2 \exp[( -E_0 + E_{\rm R}/n^2 )/\kb T]$,
which is infinite. Setting this equal to $2$ without a thorough argument is hardly justified!}.

Eqn (\ref{Gibbs}) gives, for the contribution to chemical potential from translational kinetic energy,
\be
\mu_i^*  = \left(\kb+ \frac{C_{V,i}}{N_i}\right) T - T \frac{S_i}{N_i}  .
\ee
By substituting this three times into (\ref{muERstar}) one obtains eqn (\ref{selectron}) as before.

\vspace{1ex}

{\em Justification of  $\zeta_{\rm H} \simeq 2 \zeta_{\rm p}$.}
The fact that $\zeta_{\rm H} \simeq 2 \zeta_{\rm p}$, that we used in order to obtain equation
(\ref{Selectron}), is a standard part of the derivation of the Saha equation, but ordinarily it is
justified by the use of Boltzman's statistical formula for the entropy, which we wish to avoid. To
obtain it from thermodynamics alone, one may proceed as follows.

Consider the general problem of a gas of $N$ particles with $Z$ equally likely distinct
internal states per particle, where the motion of each particle is independent of its internal
state. The main concept we need is
that such a gas has all the same thermodynamic properties as an equal mixture of $Z$
gases of $N/Z$ particles each, where each of the gases in the mixture is of the same
type as the unpolarized gas, but fully polarized, that is, with all particles in the same
internal state, and each of the $Z$ cases appears equally in the mixture.
Of course the counting of the internal states is explained by quantum theory, but all we need is
the empirical fact that if one somehow prepares the mixture just described, then the system
that results will be thermodynamically the same as the unpolarized gas.

Let $S$ be the total entropy of the gas, and $S_m$ the entropy of one of the components
in the above model. Then $S = \sum_i S_i$. Applying 
eqn (\ref{entropy}) to each term in this equation, we find
\be
N k_{\rm B} \ln \left[ \zeta \frac{V}{N} \kappa \right] 
= \sum_{i=1}^Z  N_i k_{\rm B} 
\ln \left[ \zeta_i \frac{V}{N_i} \kappa \right]    \label{Szetam}
\ee
where $N_i = N/Z$ and $\kappa = [k_{\rm B}T/(\gamma-1)]^{1/(\gamma-1)}$.
Here, $\zeta$ is the constant appearing in the formula for the entropy of the whole gas,
and $\zeta_i$ are the constants appearing in the formulae
for the entropies of the component gases.
We now claim that all the $\zeta_i$ are the same, because the motions of the particles
are independent of their internal state.
Substituting this into (\ref{Szetam}) gives $\zeta = Z \zeta_i$. It follows that,
for two gases that are identical except that one has 4 internal states per particle, the other 2,
one will find the $\zeta$ of the first will be twice that of the second. QED.

\vspace{12pt}

\bibliography{thermorefs}

\end{document}